\documentclass[a4paper,twocolumn,showpacs,nofootinbib,floatfix,superscriptaddress,prl]{revtex4-1}
\usepackage{graphicx}
\usepackage{amsfonts}
\usepackage{amsmath}
\usepackage{hyperref}
\usepackage{float}
\usepackage{relsize}

\newcommand{\be}{\begin{equation}}			
\newcommand{\ee}[1]{\label{#1} \end{equation}}		
\newcommand{\ba}{\begin{eqnarray}}			
\newcommand{\ea}[1]{\label{#1} \end{eqnarray}}		
\newcommand{\nl}{\nonumber \\}				

\newcommand{\td}[2]{ \frac{{\rm d\,} #1}{{\rm d\,} #2}}				

\begin{document}




\title{
{\bf Derivation of Tsallis  Entropy }{\bf and the Quark-Gluon-Plasma Temperature}  
}

\author{T.S.\ Bir\'o, G.G.\ Barnaf\"oldi \, and \, P.\ V\'an}
\affiliation{Wigner Research Centre for Physics of the HAS, 
H-1525 Budapest, P.O.Box 49, Hungary}

\begin{abstract}
We {derive} Tsallis entropy, $S_q$, from universal thermostat independence
and obtain the functional form of the corresponding generalized entropy-probability relation.
Our result for finite thermostats {interprets thermodynamically} the subsystem temperature, $T_1$, and
the index $q$ in terms of the temperature, $T$, entropy, $S$, and heat capacity, $C$ 
of the reservoir as $T_1=T\exp(-S/C)$ and $q=1-1/C$.
In the infinite $C$ limit, irrespective to the value of S, the Boltzmann\,--\,Gibbs
approach is fully recovered.
We apply this framework for the experimental determination of the original
temperature of a finite thermostat, $T$, from the analysis of hadron spectra
produced in high energy collisions, by analyzing frequently considered
simple models of the quark-gluon plasma.
\end{abstract}

\date{September 19, 2012.} 


\maketitle


A nonlinear entropy formula  has been suggested by R\'enyi
long ago and been applied to several areas in physics
\cite{Renyi:1970,Lenzi:2000gj,Bashkirov:2003,Jizba:2003gx,Bialas:2008zz,Parvan:2010,Klebanov:2011uf}. 
Another formula, the Tsallis entropy, has more recently been promoted as the keystone for a 
generalized thermodynamics, treating correlated physical systems 
\cite{Tsallis:1988,Tsallis:1998,Tsallis:1999,TsallisBook}. 
A respectable amount of papers applying this idea to one or the other area in physics appeared 
\cite{TsallisLink,Plastino:1999wk,Abe:1999wk,Chen:2002hx,Kaniadakis:2004rj,Mathai:2006gh}. 
Since from this entropy the canonical energy distribution
is power-law tailed in place of the Boltzmann\,--\,Gibbs exponential, 
numerous high-energy distributions have been fitted using the Tsallis formula 
\cite{Biro:2005uv,Biro:2008er,Biro:2008hz,Urmossy:2011xk,CMSfit,Wilk:2009aa,Wilk:JPG39,Wilk:2012,Cleymans:2008mt,Cleymans:2012}.
Its independence from the thermostat and the
thermodynamical foundation behind the use of such a formula are interesting questions.

In our earlier works we investigated some general mathematical properties of alternative
entropy formulas via their pairwise composition rules, and established that a
scaled repetition of an arbitrary composition rule leads to an associative asymptotic
composition rule of large subsystems \cite{AsympRules}. All such rules are uniquely defined by a strict
monotonic function, their formal logarithm. Recently we have also observed that -- in connection to
the zeroth law of the thermodynamics -- the factorizability condition on the common entropy
maximum \cite{BiroBook} allows only for such rules \cite{ZerothLaw}.
We seek in this paper for the thermodynamical meaning of the $q$ parameter 
generalizing the classical entropy formula, valid for $q=1$.
Some $q \ne 1$ parameter was calculated theoretically 
\cite{OpticalLatticeTheory,OpticalLatticeExperiment,CentralCharge}.


We found the thermodynamical interpretation of 
the entropy formula and its parameters on the analysis of the two-body thermodynamics
of a single observed subsystem and a reservoir.
For finite systems the microcanonical approach is the key to the physical interpretation.
In the classical treatment subleading terms in a finite-energy expansion
of the microcanonical entropy maximum are usually ignored, the reservoir is treated as constant
in the canonical limit. A notable exception is the analysis of statistical fluctuations
and their scaling in the thermodynamical limit 
\cite{VolumeFluct,BoseFermiFluct,ExactChargeFluct,MicroFluct,CanonFluct,Torrieri2010,Wilk:TempVolFluct}.
We consider the correlation between subsystem and reservoir 
induced by the conservation of total energy while maximizing a monotonic function
of the Boltzmann\,--\,Gibbs entropy, $L(S)$.  We seek for that very function, $L$, 
which counteracts finite size effects beyond the usual linear term,
$- \beta E_i$, in the Taylor-expansion of the $L(S)={{\rm max}}$ principle.

We discuss now the thermal equilibrium of two systems, one with energy $E_1$ (subsystem) and the other
with energy $E-E_1$ (reservoir), while their respective entropy contributions are
combined by the general rule satisfying
\be
 L(S_{12}) = L(S_1) + L(S_2).
\ee{S_RULE} 
Here we do not assume that the deviation from the simple additive rule would be small.
For the sake of simplicity we consider here homogeneous rules, relevant for the cases
when subsystem and reservoir are composed from the same matter (for details see Ref.\cite{ZerothLaw}).
The microcanonical condition for a maximal entropy state then
defines the thermodynamical inverse temperature, requiring
\be
 L(S(E_1)) + L(S(E-E_1)) = {\rm max.} 
\ee{ENTR_MAX} 
Varying the subsystem energy, $E_1$, while keeping the total
energy $E$ fixed, we describe the thermal contact between subsystem and reservoir. 
This means that the derivative with respect to $E_1$ of the above expression (\ref{ENTR_MAX}) vanishes.
Owing to the two $E_1$-dependent contributions, it is equivalent to the statement that
\be
 \beta_1 = L'(S(E_1)) \cdot S'(E_1)  = L'(S(E-E_1)) \cdot S'(E-E_1).
\ee{ENT_DERIV}
This equality, when taken in the $E \gg E_1$ limit, usually defines the canonical approach.
Now we would like to take into account effects to higher order in $E_1/E$, and require that 
their leading term vanishes on the right hand side. 
The reservoir's entropy on the right hand side is Taylor-expanded:
$S(E-E_1)= S(E) - S'(E) \cdot E_1 +\ldots$.
Collecting the coefficients of $E_1$ we arrive at
\ba
& &   \beta_1 =  L'(S(E)) \cdot S'(E)  
\nl
& & - \left[S'(E)^2L''(S(E))+S''(E)L'(S(E)) \right] E_1 + \ldots 
\ea{BETA_LINEAR}
Here the first term on the right hand side is the familiar canonical ($E_1$-independent) Lagrange multiplier,
\be
 \beta = L'(S(E)) \cdot S'(E) = L'(S) \cdot \frac{1}{T},
\ee{DEFINE_BETA}
constituting the $\beta_1=\beta$ relation.
Our key addition to the usual treatment is to 
{require that the coefficient of the linear term  in eq.(\ref{BETA_LINEAR}) vanishes}: 
This is a constraint for the $L(S)$ function in general.
Obviously, without considering $L(S)$, the whole coefficient consisted only of $S''(E)$
as in the traditional approach, and nothing further could be done.
We obtain the following condition:
\be
 \frac{L''(S)}{L'(S)} = - \frac{S''(E)}{S'(E)^2}.
\ee{FORMLOG_DIFFEQ}

Since the left hand side of eq.(\ref{FORMLOG_DIFFEQ}) is a function of $S$, 
while the right hand side is a function of $E$, the left hand side must be treated as an $S$-independent constant
by solving eq.(\ref{FORMLOG_DIFFEQ}) for $L(S)$. 
This {\em Universal Thermostat Independence} (UTI) reads as
\be
 \frac{L''(S)}{L'(S)} = a. 
\ee{FORMLOG_EQ_SPECIAL}
The particular solution with $L'(0)=1$ and $L(0)=0$ is given by
\be
 L(S) = \frac{e^{aS}-1}{a}.
\ee{FORMLOG_SPECIAL}
The derivatives of the $S(E)$ equation of state do have physical meaning:
$S'(E)=1/T$ and $S''(E)=-1/CT^2$ are related to the traditional temperature and
heat capacity of the reservoir. By using this we obtain
\be
 a = {1}/{C}.
\ee{NON_ADDITIVITY_PARAMETER}
{\em The non-additivity parameter is simply the inverse heat capacity} of the reservoir.
For $C\rightarrow\infty$ one has $a\rightarrow 0$ and $L(S)\rightarrow S$, so the Boltzmann\,--\,Gibbs formula
is included by this limit.
A connection between Tsallis entropy and constant heat capacity
of the reservoir has been observed years ago \cite{Almeida:2001,Rybczynski:2004gs}.
Our result shows the background for this observation.
The philosophy behind our approach is first to decide on the entropy formula by choosing $L(S)$
generally and then to solve the maximization problem in terms of subsystem energies
and corresponding probabilities. 

The knowledge gained from this analysis now will be generalized.
Analog to a Gibbs ensemble, we extend a sum of two to a weighted sum of many. 
In this way the result of the two-body analysis in the form
$L(S_1)$ generalizes the classical entropy formula, $S=-\sum_i P_i \ln P_i$, to
\be
 L(S) = \sum_i P_i L(-\ln P_i).
\ee{GENERAL_ENTROPY} 
This {\em L-additive} form of the generally non-additive entropy
leads to the Tsallis entropy formula when applying eq.(\ref{FORMLOG_SPECIAL}).
In this way one obtains
\be
 L\left(S(E_1)\right) - \beta E_1 = \frac{1}{a}\left(e^{aS(E_1)}-1 \right) - \beta E_1 = {{\rm max.}}
\ee{CANONICAL1}
Here the coefficient of the second order correction, ${\cal O}\left(E_1^2/E^2\right)$, 
vanishes for $a=1/C(E)$. By this we are led
to the following entropy expression 
\be
 L(S(E_1)) = L(-\ln P_1) = \frac{1}{a} \left( P_1^{-a} - 1 \right).
\ee{MAXPROB_ENTROPY}
Our result applied to a Gibbs-ensemble with the relative occurrence frequency $P_i$ of
states with energy $E_i$, hence reads as
\be
 \sum_i P_i L(-\ln P_i)  - \beta \sum_i P_i E_i  - \alpha \sum_i P_i  = {{\rm max.}}
\ee{CORRECTED_CANONICAL}
Substituting eq.(\ref{MAXPROB_ENTROPY}) we finally arrive at 
\be
 \frac{1}{a} \sum_i \left(P_i^{1-a} - P_i \right) -  \beta \sum_i P_i E_i  - \alpha \sum_i P_i  = {{\rm max.}}
\ee{TSAL_AS_COR_CANON}
With the widespread notation $q=1-a$ one obtains the Tsallis entropy formula 
\be
 S_{{\rm Tsallis}} := L(S) = \frac{1}{q-1} \sum_i (P_i-P_i^q). 
\ee{SIGMA_TSALLIS}
It is suggestive to consider its inverse function, $L^{-1}$ according to eq.(\ref{GENERAL_ENTROPY}).
This delivers the R\'enyi entropy:
\be
 S_{{\rm R\acute{e}nyi}} := S = \frac{1}{1-q} \ln \sum_i P_i^q. 
\ee{RENYI_AS_FINAL_CONCLUSION}
Now the parameters $\beta$ and $a$, defined in eqs.(\ref{DEFINE_BETA}) and (\ref{NON_ADDITIVITY_PARAMETER}), 
are set by the physics of the finite-energy reservoir.
The sign of the heat capacity, $C$, determines whether the $q$ is smaller or larger than one.
It may possibly carry an interesting message for the description of gravitating systems,
with negative heat capacity.

We proceed by noting that maximizing  $S_{{\rm Tsallis}}$  with respect to
the $P_i$ weights of system instances with energy $E_i$ one obtains the canonical
cut power-law distribution of energies:
\be
 P_i =  \left(Z^{1-q} + (1-q) \frac{\beta}{q} E_i \right)^{\frac{1}{q-1}}.
\ee{CANON_ENERG_DISTRIB}
Using eqs.(\ref{DEFINE_BETA}), (\ref{FORMLOG_SPECIAL}), and (\ref{NON_ADDITIVITY_PARAMETER}) 
we rewrite this in the equivalent form,
\be
 P_i =  \frac{1}{Z} \left( 1 + \frac{Z^{-1/C} \, e^{S/C}}{C-1} \frac{E_i}{T} \right)^{-C},
\ee{P_CANON_EQUIV}
expressing the energy distribution in terms of the temperature, $T$, entropy, $S$ and heat capacity,
$C$ of the ideal reservoir. The partition sum $Z$, obtained from normalization, is related
to the Tsallis-entropy, $L(S_1)$, and energy, $E_1$, of the subsystem via its deformed logarithm:
\be
 \ln_q Z := C\left( Z^{1/C}-1\right) = L(S_1) - \frac{1}{1-1/C} \beta E_1.
\ee{ZPART}
In the infinite heat capacity limit, irrespective to the value of $S$, 
formula (\ref{P_CANON_EQUIV}) recovers the exponential distribution.
The inverse logarithmic slope of the energy distribution, derived from it, is linear:
\be
 T_{{\rm slope}}(E_i) = \left(-\td{}{E_i} \ln P_i \right)^{-1} = T_0 + E_i/C,
\ee{T_SLOPE}
with $T_0 = T  e^{-S/C} Z^{1/C} (1-1/C)$.
One concludes that the generalized entropy formula leads to a cut power-law
energy distribution, based on a finite heat capacity reservoir. 
As such, it is a better approximation to the microcanonical distribution, 
than the canonical exponential.

\begin{figure}
\begin{center}
        \includegraphics[width=0.45\textwidth]{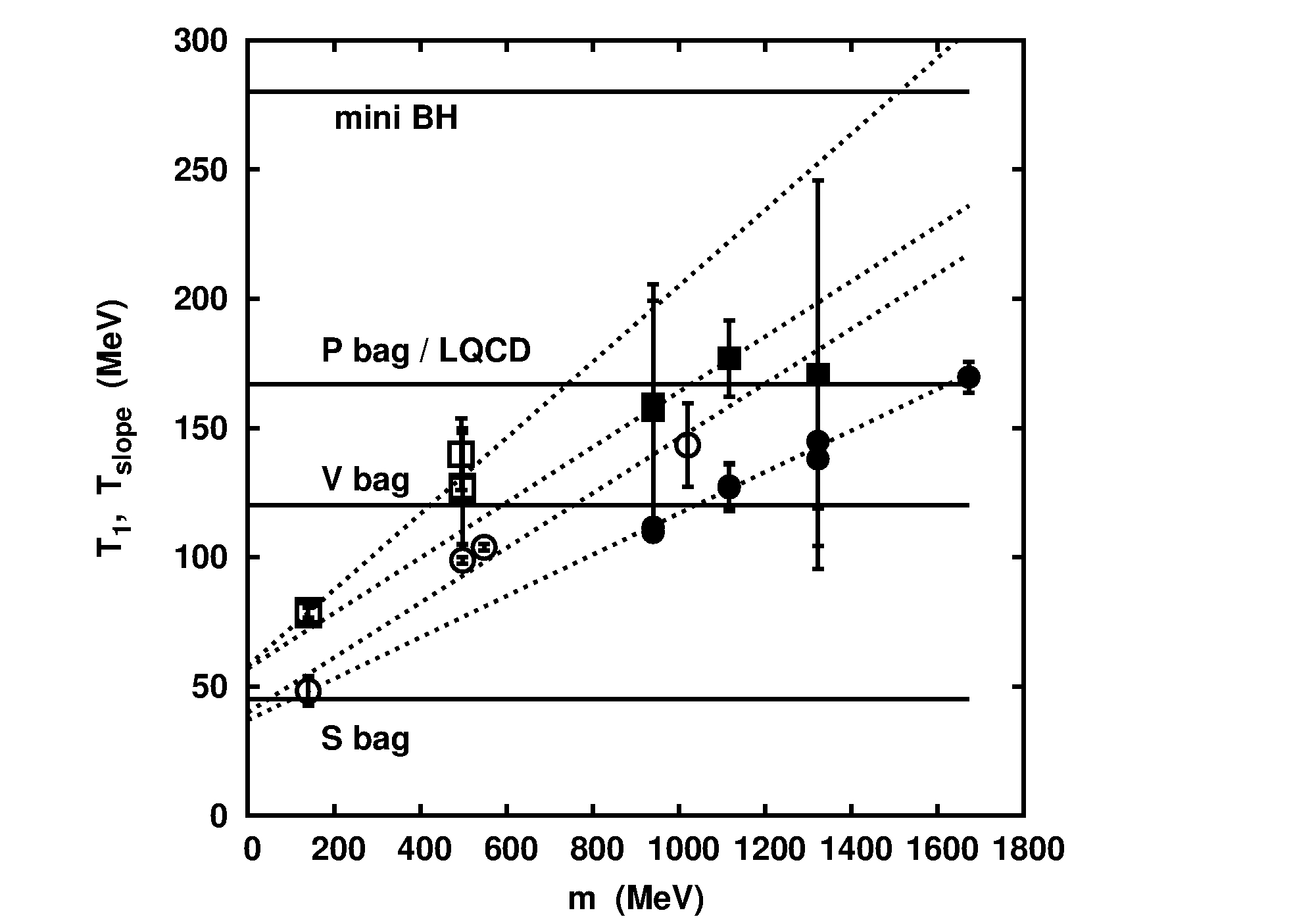}
\end{center}
\caption{ \label{Fig1}
 Extrapolated inverse slopes from RHIC $AuAu$ data at $200$ GeV (circles),
 LHC $pp$ data at $900$ GeV (boxes) and for different theoretical models of
 a QGP-thermostat with temperature $T=167$ MeV (horizontal lines).
 Open symbols correspond to mesons, filled symbols to baryons. The skew lines
 indicate the respective valence quark assumptions: 
 $T_{{\rm slope}}^{{\rm meson}}(p_T=0) = T_0 + m_i/2C$ and
 $T_{{\rm slope}}^{{\rm baryon}}(p_T=0) = T_0 + m_i/3C$, for the LHC (upper two lines) and
 for RHIC (lower two lines).
}

\end{figure}

We demonstrate the usefulness of the above general results on 
the example of the thermal model to heavy-ion collisions.
Experimental data from RHIC $AuAu$ collision at $200$ GeV deliver different $T_{{\rm slope}}$-s
extrapolated to $p_T=0$ for different hadrons \cite{Biro:2008er,BiroUrmSchram}. Considering that the energy at
zero momentum is the rest mass in $c=1$ units, 
a linear trend shows in the $T_{{\rm slope}}(m_i)$ values, as seen in
Fig.\ref{Fig1}. The open circles correspond to mesons, the filled ones to baryons
in this figure. The steepness for mesons and baryons seem to be in the proportion
$2:3$ suggesting a quark coalescence hadronization picture, compatible with
the factorization assumption $P_{{\rm hadron}}(E)=P_i^K(E/K)$ with $K=2$ and
$K=3$ for mesons and baryons, respectively. This scaling is acceptable
and leads to $T_{{\rm slope}}^{{\rm hadron}}(E)=T_{{\rm slope}}^{{\rm quark}}(E/K)$, with the 
common $T_0 \approx 48$ MeV intersect in the formula (\ref{T_SLOPE}). The valence quark matter
heat capacity at RHIC $AuAu$ collision tends to be $C\approx 4.5$. 
Similar trends can be extracted from the analysis of fits
to the ALICE data in $900$ GeV $pp$ collisions done in \cite{CleymansWorku}: the values for the
Tsallis-slope parameters, $T_0$, are much lower than the canonical QCD phase transition
temperature. Here we re-plotted the tabulated values given in \cite{CleymansWorku}
using the coalescence quark assumption, denoting mesons by open square boxes
while baryons by filled boxes. We note, however, that these fits
were performed in the very low $p_T$ range ( $p_T < 2.5 $ GeV/c) only, therefore
the uncertainty of the fitted parameters is large.

In order to interpret this surprisingly low value for $T_0$, 
we have to consider physical models of a finite thermostat, and calculate
$T_1=1/\beta_1=Te^{-S/C}$, since $\lim_{C\rightarrow\infty}\limits T_0 = T_1$ for
small subsystems in large reservoirs.
First we study the Stefan\,--\,Boltzmann formula supplemented with a bag constant,
$E/V=\sigma T^4 + B$ in a volume $V$. 
Since the pressure is given by $p=\frac{1}{3}\sigma T^4-B$, the entropy 
is $S=\frac{4}{3}\sigma VT^3$. 
The heat capacity is the derivative of the energy with respect to temperature,
\be
 C = \td{E}{T} = 4 \sigma VT^3 + \left(\sigma T^4 + B \right) \td{V}{T}.
\ee{C_HEAT_QGP}
At constant volume, $V$, this gives $C_V=4\sigma VT^3=3S$
and $T_{1V}=Te^{-1/3}$.
At constant pressure the temperature cannot change in this model, so $C_p=\infty$
and $T_{1P}=T$. 
Furthermore, considering an adiabatically expanding reservoir, a more realistic
scenario in high-energy experiments, one deals with the heat capacity at constant
entropy, $C_S=3S(1-T_*^4/T^4)/4$, with $T_*$ being the temperature where the pressure
vanishes. In this case $C_S \le 3S/4$ and $T_{1S} \le Te^{-4/3}$ is the theoretical
prediction.

Figure \ref{Fig1} presents the inverse logarithmic slope, $T_{{\rm slope}}$,
as a function of hadron masses and $T_1$-lines  for different physical models
of the thermostat.
Besides the three above described bag-model approaches we also indicate 
the classical Schwarzschild  black hole, having $C=-2S$ and $T_1=Te^{1/2}$.
One inspects that this possibility is far from all experimental observations.
 
We note that theoretically a really constant
heat capacity, $C_0$, stems from the equation of state $S(E)=C_0 \ln(1+\frac{E}{C_0T_0})$. The latter is
a good ansatz for an effective equation of state of classical non-abelian gauge field
systems on the lattice \cite{Chaos} and represents the high-$E$ limit of Planck's 
$S''(E)$-formula for thermal radiation.

Considering heat capacity in the above scenarios and a
standard numerical value of $T\approx 167$ MeV for the reservoir temperature, 
conjectured for the QGP at hadronization phenomenology and determined by lattice QCD calculations,
one obtains $T_{1P}=T=167$ MeV, $T_{1V}=Te^{-1/3}\approx 120$ MeV 
and $T_{1S} \le Te^{-4/3} \approx 45$ MeV 
characterizing the Tsallis-distribution of valence quarks. 

The conjecture that in heavy ion collisions a statistical power-law energy distribution
due to finite phase space availability corrections to the traditional canonical
distribution may appear is further supported by the observation that the measure
of non-additivity, $a=1/C$, expressed by the inverse power in the fitted power-law tail,
is reduced for increasing participant number \cite{RHIC_Participant}. 
The fitted power $C$ is also tendentiously smaller in $e^+e^-$ or $pp$ than in heavy ion collisions \cite{Urmossy:2011xk}. 
Finally we realize that only the adiabatic scenario for the quark matter thermostat
leads to $T_0$ values near to the ones extracted from experimental
analysis by the coalescence assumption, $T_{1S}\approx T_0 \approx 45 - 55$ MeV.


In conclusion the Tsallis entropy formula is derived as the consequence of the following
requirement: we seek for that non-additive entropy composition rule which cancels
linearly energy-dependent corrections due to the finite $E-E_1$
energy in the reservoir to a subsystem's thermodynamical inverse temperature.
This determines the composition rule and the entropy formula 
uniquely turns out to be the Tsallis entropy. 
This derivation explains the particular functional form of the
Tsallis and R\'enyi formulas as generalized entropy expressions satisfying the UTI principle.
With regard to the physical interpretation we have obtained $ q = 1 - 1/C$,
with $C$ being the heat capacity of the total system with the conserved energy $E$.
The canonical temperature of the subsystem becomes
$ T_1 = e^{-S/C} \, T$
with $T(E)$ and $S(E)$ being the traditional temperature and entropy of the finite reservoir, respectively.
A preliminary analysis of experimental data on particle production seems to
be sensitive to different physical assumptions about a QGP thermostat.
Here the isentropic scenario performs best.

\vspace{2mm}
{\em Acknowledgment: \quad}
Discussions with Prof. C. Tsallis are gratefully acknowledged. We thank
K. \"Urm\"ossy for providing slope values on RHIC $AuAu$ data at $200$ GeV.
This work was supported by Hungarian OTKA grants NK778816, NK106119,
H07-C 74164, K68108, K81161, K104260, NIH TET\_10-1\_2011-0061, and ZA-15/2009.
Author GGB also thanks the J\'anos Bolyai Research Scholarship of the
HAS.



\end{document}